\newcommand{\la}{\langle}
\newcommand{\ra}{\rangle}

\newcommand{\beq}{\begin{eqnarray}}
\newcommand{\eeq}{\end{eqnarray}}

\newcommand{\bff}{\mbox{{\boldmath $f$}}}

\newcommand{\bfu}{{\bf u}}

\newcommand{\bfv}{\mbox{{\boldmath $v$}}}
\newcommand{\bfr}{\mbox{{\boldmath $r$}}}
\newcommand{\bfj}{\mbox{{\boldmath $j$}}}

\newcommand{\tf}{\tilde{f}}
\newcommand{\tbff}{\tilde{\bff}}
\newcommand{\tbfr}{\bar{\bfr}}

\newcommand{\bfJ}{\mbox{{\boldmath $J$}}}

\newcommand{\bfpsi}{\mbox{{\boldmath $\psi$}}}
\newcommand{\bfphi}{\mbox{{\boldmath $\varphi$}}}

\newcommand{\eps}{\epsilon}

\renewcommand{\d}{\partial}

\newcommand{\btem}{\bibitem}

\newcommand{\NGo}{N. \ Goldenfeld}
\newcommand{\YO}{Y.\ Oono}
\documentclass[fleqn,11pt,twocolumn]{article}

\usepackage[dvips]{graphicx}
\title{%
The Renormalization-group Method Applied to Non-equilibrium
Dynamics
\\
\vspace{3mm}
\normalsize
\begin{center}
\vspace{-.2cm}
Teiji Kunihiro\\
{\it Yukawa Institute for Theoretical Physics, Kyoto University,
 Kyoto, 606-0952, Japan}
\end{center}
}
\vspace{-.2cm}
\date{\empty}
\begin{document}
\maketitle
\begin{abstract}
We review the renormalization group method applied
to non-equilibrium dynamics by tracing the way how the hydrodynamic
equations can be derived as reduced dynamics of the Boltzmann equation
as a typical example. 
\end{abstract}

\section{Introduction}
The theory of the nonequilibrium dynamics
 may be regarded as a collection of the theory 
how to reduce  the dynamics 
of many-body systems to ones with fewer variables
\cite{kuramoto}.
In fact, the  Bogoliubov-Born-Green-Kirkwood-Yvon
hierarchy\cite{reichl}  can be reduced to the time-{\em ir}reversible
 Boltzmann equation\cite{boltzmann}, which is 
 given solely in terms of the single-particle
 distribution function for dilute gas systems\cite{bogoliubov}.
The derivation of Boltzmann equation by Bogoliubov\cite{bogoliubov}
 shows that  the dilute-gas dynamics as a {\em dynamical system}
with many-degrees of freedom 
has an {\em attractive manifold} \cite{holmes} spanned by 
 the one-particle distribution function,
 which is also an {\em invariant manifold} \cite{holmes}. 
Boltzmann equation in turn can be further
  reduced to the hydrodynamic equation 
(Euler or Navier-Stokes equation\cite{landau})
   by a perturbation theory like Chapman-Enskog 
  method\cite{chapman} or Bogoliubov's 
method\cite{bogoliubov,krylov}.
Langevin equation\cite{fp} which may be time-irreversible can be
reduced to the time-irreversible Fokker-Planck equation
with a longer time scale than the scale in Langevin 
equation\cite{fp}.

Two basic ingredients are 
commonly seen in the reduction of dynamics, which are interrelated  
with  but relatively independent of each other:
(i) The reduced dynamics is characterized 
with a longer time scale than that appearing in the
original (microscopic) evolution equation.
(ii) The reduced dynamics is described by
 a time-irreversible equation even when the
 original microscopic equation is 
time-reversible\cite{lebowitz,kawasaki}.
The fundamental problem in the theory of deriving kinetic or transport
 equations is to clarify the mechanism of and to implement the above two
basic ingredients.

In a few years ago, we  showed\cite{aop02} that the so called the
renormalization-group method\cite{RG,shirkov,cgo}, 
which is formulated in terms of the
classical theory of envelopes\cite{kuni95}, gives a unified theory for the
reduction of the dynamics and can be used to derive the various
transport equations in a transparent way; we have also elucidated that 
the underlying mathematics of the reduction is an explicit
construction of the invariant manifold\cite{aop02,efk},
 a notion in dynamical system\cite{holmes,krylov}.

In this report, we pick up the problem to derive the hydrodynamics
equations from the Boltzmann equation as a typical problem of the
reduction appearing in the field of non-equilibrium dynamics.
Some remarks will be also given 
on applications of the method to other problems such as
the reduction of the dynamics of Fokker-Planck equation and also
on an extension to the relativistic case.

\section{Derivation of hydrodynamic equations from Boltzmann equation}


In this section, 
we apply the RG method formulated in \cite{kuni95,efk}  to 
 derive Euler and Navier-Stokes
equations, successively from the Boltzmann equation\cite{aop02}.

\subsection{Basics of Boltzmann equation}

Boltzmann equation\cite{boltzmann,resibois} is an evolution
 equation of the one-particle distribution function
$f(\bfr, \bfv, t)$ in the phase space, and reads
\beq
\frac{\d f}{\d t}+\bfv\cdot \frac{\d f}{\d \bfr}=I[f].
\eeq
Here the left-hand side
describes the change due to the canonical equation of 
motion while
 the right-hand side the change due to collisions;
\beq
I[f]&=&\int d\bfv_1\int d\bfv'\int d\bfv_1'w(\bfv\, \bfv_1
 \vert \bfv'\, \bfv_1')
      \nonumber \\ 
 \quad & & \times \biggl\{f(\bfr, \bfv', t)f(\bfr, \bfv_1',
 t)\nonumber \\
 &-& f(\bfr, \bfv, t)f(\bfr, \bfv_1, t)\biggl\},
\eeq
which is called the collision integral.
The transition probability
$w(\bfv\, \bfv_1 \vert \bfv'\, \bfv_1')$ has the following
 symmetry due to the time-reversal invariance of the
microscopic equation of motion;
$w(\bfv\, \bfv_1 \vert \bfv'\, \bfv_1')=w(\bfv'\, \bfv_1' \vert \bfv\, \bfv_1).$
Furthermore, the invariance under the particle-interchange
 implies the following equality;
$w(\bfv\, \bfv_1 \vert \bfv'\, \bfv_1')=w(\bfv_1\, \bfv \vert
\bfv_1'\, \bfv')$
$=w(\bfv_1'\, \bfv' \vert \bfv_1\, \bfv)$.
%


We say that the function 
$\varphi (\bfv)$ is a {\em collision invariant} if 
it satisfies the following equation;
\beq \int d\bfv\varphi(\bfv)I[f]=0. 
\eeq
Owing to the conservation  of the particle number,
the total momentum and the total kinetic energy during the
collisions, 
the quantities given by linear combination of the following
five quantities are collision invariant:
$1, \bfv $ and $v^2$.

For a collision invariant $\varphi (\bfv)$,
 we can define the density $n_{\varphi}$ and the current
 $\bfj_{\varphi}$ as follows;
\beq 
n_{\varphi}&=&\int d\bfv
\varphi (\bfv)f(\bfr, \bfv, t),\nonumber \\
 \bfj_{\varphi}&=&\int d\bfv \bfv
\varphi (\bfv)f(\bfr, \bfv, t),
\eeq
which satisfy the continuity or balance equation;
\beq
\label{balance}
\d_t n_{\varphi}+\nabla\cdot \bfj _{\varphi}=0.
\eeq
Thus we have formally the 
hydrodynamic equations as the balance equations for 
the conservation of the particle number,
 total momentum and kinetic energy.
These equations are, however,  formal ones because
the distribution function $f$ is not yet solved:
The solution obtained from Boltzmann equation 
will give  the explicit forms of the internal energy,
 the transport coefficients and so on.

The $H$ function is defined as follows:
\beq
H(\bfr, t)=\int d \bfv \, f(\bfr , \bfv , t)\big(\ln f(\bfr ,\bfv , t) -1\big).
\eeq
For equilibrium states, the $H$ function
 is equal to the entropy $S$ with the sign changed.
Defining the corresponding current by
\beq
\bfJ _H(\bfr , t)=\int d \bfv \, \bfv\,
 f(\bfr , \bfv , t)\big(\ln f(\bfr ,\bfv , t) -1\big),
\eeq
one has the balance equation;
\beq
\frac{\d H}{\d t}+\nabla\cdot \bfJ _H=\int d \bfv \, I[f]\ln f.
\eeq
This shows that when $\ln f$ is a collision invariant,
the $H$ function is conserved.
In fact, one can show that  this condition is satisfied when
$f(\bfr , \bfv , t)$ is a local equilibrium distribution
 function as given by (\ref{lmaxwell}) below.

\subsection{Application of 
 the RG method to derive hydrodynamic equations}

To make it clear that the following discussion fits
 to  the general formulation given in \cite{efk},
 we discretize the argument $\bfv $ as 
$\bfv \rightarrow \bfv _i$\cite{kuramoto}:
 Discriminating the arguments $(\bfr , t)$ and $\bfv _i$ in
$f(\bfr, \bfv _i, t)$, we 
indicate $\bfv _i$ as a subscript $i$ for the 
distribution function;
$
f(\bfr, \bfv _i, t)=f_{i}(\bfr , t)\equiv (\bff (\bfr , t))_i.
$
Then Boltzmann equation now reads
\beq
\label{dbol-0}
\frac{\d f_{i}}{\d t} = \hat{I}[\bff]_i
- {\bfv}_i \cdot  \frac{\d f_{i}}{\d \bfr},
\eeq
where 
\beq
\hat{I}[\bff]_i = 
\sum_{j,k,l} w( {\bfv}_i{\bfv}_j\vert {\bfv}_k{\bfv}_l)
( f_kf_l - f_if_j )({\bfr}, t).
\eeq

Now let us consider a situation where
 the fluid motion is slow with long wave-lengths so that
\beq
\bfv _i\cdot \frac{\d f_i}{\d \bfr}=O(\eps), 
\eeq
where  $\eps $ is a small quantity,$ \vert \eps\vert <1.$
To take into account  the smallness of $\eps $ in the 
following calculations formally,
 let us introduce the scaled coordinate $\tbfr$ defined by
$
\tbfr =\eps \bfr$ and
$\d/\d \bfr=\eps \cdot \d/\d \tbfr$. 
Then (\ref{dbol-0}) reads
\beq
\label{dbol-1}
\frac{\d f_{i}}{\d t} = \hat{I}[\bff]_i
- \eps {\bfv}_i \cdot  \frac{\d f_i}{\d \tbfr},
\eeq
which has a form to which the 
 perturbation theory given in \cite{efk} is naturally applicable.

In accordance with the general formulation given in \cite{efk},
 we first expand the solution as follows;
$
f_i( {\tbfr},t ) = 
f_i^{(0)}( {\tbfr},t ) + \eps f_i^{(1)}( {\tbfr},t )+\cdots .
$
Let $\tf _i(\tbfr , t; t_0))$ be a solution around
$t\sim t_0$ given by a perturbation theory with  
$f_i(\tbfr , t_0)$ being the initial value
 at $t=t_0$;
\beq
\tf _i(\tbfr , t=t_0; t_0)=f_i(\tbfr , t_0).
\eeq
We expand $\tf _i(\tbfr , t; t_0)$ as 
\beq
\tf_i( {\tbfr},t; t_0 )& =& \tf_i^{(0)}( {\tbfr},t ; t_0 ) +
 \eps \tf_i^{(1)}( {\tbfr}, t, t_0 )\nonumber\\
 &+& \cdots, 
\eeq
and the respective initial condition
 is set up as follows,
\beq
\tf _i^{(l)}(\tbfr , t=t_0; t_0)=f_i^{(l)}(\tbfr , t_0), 
 \quad (l=0, 1, 2 ...).
\eeq

The $0$-th order equation reads
\beq
\frac{\d \tf _i^{(0)}}{\d t} = (I[{\tbff} ^{(0)}])_i.
\eeq
Now we are interested in 
 the slow motion which may be achieved 
 asymptotically as $t\rightarrow \infty$.
Therefore we take the following stationary 
 solution,
\beq
\frac{\d \tf _i^{(0)}}{\d t} = 0,
\eeq
which is  a {\bf fixed point} of the 
 equation satisfying
\beq
\label{bol-fix}
(\hat{I}[{\tbff} ^{(0)}])_i=0,
\eeq
 for arbitrary $\tbfr$.
Notice that (\ref{bol-fix}) shows that 
the distribution function ${\tbff} ^{(0)}$ is a
 function of collision invariants.
Such a distribution function is a local equilibrium 
distribution function or Maxwellian;
\beq
\label{lmaxwell}
\tf_i^{(0)}(\tbfr , t; t_0 )
 &=& n(\tbfr , t_0)
\biggl(
\frac{m}{2 \pi k_B T(\tbfr , t_0)}
\biggl)^{3/2}\nonumber \\
&\times&\exp \biggl[
- \ \frac{m \vert {\bfv}_i - {\bfu}(\tbfr ,t_0) \vert ^2}
{2 \pi k_B T(\tbfr ,t_0)}
\biggl].
\eeq
Here,
the local density $n$, local temperature $T$, local
 flux ${\bfu}$ are all dependent on
 the initial time $t_0$ and the space coordinate $\tbfr$
 but independent of time $t$.

The first-order equation reads
\beq
\biggl(
(\frac{\d}{\d t} - A){\tf}^{(1)}
\biggl)_i 
= - {\bfv}_i \cdot \frac{\d \tf_i^{(0)}}{\d {\tbfr}}.
\eeq
Here the linear operator $A$ is defined by
\beq
\biggl[\hat{I}'[\tbff ^{(0)}]{\tbff}^{(1)}\biggl]_i 
&=&
\sum_{j=1}^{\infty} \frac{\d I}{\d \tf _j}
\biggl\vert _{{\tbff}={\tbff}^0}\cdot \tf _j^{(1)} \nonumber \\
&\equiv& ( A{\tbff}^{(1)} )_i.
\eeq
Defining the inner product between $\bfphi$ and $\bfpsi$ by
\beq
\la \bfphi, \bfpsi \ra =  \int d{\bfv} \varphi \psi, 
\eeq
one can show that $A$ is self-adjoint;
\beq
\la \varphi, A\psi\ra =\la A\varphi, \psi\ra.
\eeq
One can further show that 
the five invariants $m, \bfv , \frac{m}{2}{v}^2$
span the kernel of $A$\cite{resibois};
$
{\rm KerA} = \{ m,{\bfv},\frac{m}{2}{\bfv}^2 \}.
$
The other eigenvalues are found to be 
negative because one can show
\beq
\la \varphi ,A \varphi \ra \quad  \le \quad   0.
\eeq

We write the projection operator 
 to the kernel as P and 
 define Q $= 1 -$ P.

Applying the general formulation in \cite{efk},
one can readily obtain the first-order solution,
\beq
\label{firstsol}
{\tbff^{(1)}} = 
- (t-t_0){\rm P}  {\bfv} \cdot 
\frac{\d \tbff ^{(0)}}{\d {\tbfr}}
+  A^{-1}{\rm Q}{\bfv} \cdot \frac{\d \tbff ^{(0)}}{\d {\tbfr}} .
\eeq
The perturbative solution up to the $\eps$ order is 
found to be 
\beq
{\tbff}(\tbfr , t, t_0)
&=& {\tbff}^{(0)}(\tbfr ,  t, t_0)
 + \eps [
- (t-t_0){\rm P}  {\bfv} \cdot 
\frac{\d \tbff ^{(0)}}{\d {\tbfr}}\nonumber \\
 & &+  A^{-1}{\rm Q}{\bfv} \cdot \frac{\d \tbff ^{(0)}}{\d {\tbfr}}
].
\eeq
Notice the appearance of a secular term.

If one stops the approximation and apply the
 RG equation\cite{kuni95},
\beq
{\d \tbff}/{\d t_0}\vert _{t_0 = t} = {\bf 0},
\eeq
 one has
\beq
\label{master:0}
 \frac{\d \tbff ^{(0)}}{\d t}  + 
\eps {\rm P} {\bfv} \cdot \frac{\d \tbff ^{(0)}}{\d \tbfr}={\bf 0}.
\eeq
This is a master equation from which equations governing
 the time evolution of $n(\bfr , t)$, $\bfu (\bfr , t)$
and $T(\bfr , t)$ in $\tbff ^{(0)}$,
 i.e., fluid dynamical equations:
In fact, taking an inner product between $m$, $m\bfv$ and 
$mv^2/2$ with this equation, one has fluid dynamical
 equations as given below.
Notice that
$f$ is now explicitly solved and 
 the energy density $e$,
 the pressure tensor $P_{ij}$ and the heat flux 
$Q_i$ are given as follows;
\beq
e(\bfr, t)&=&\int d \bfv 
\frac{m}{2}\vert \bfv - \bfu \vert ^2 f^{(0)}
(\bfr, \bfv, t)\nonumber \\
 &=&\frac{3}{2}k_BT(\bfr , t), \\
P_{ij}(\bfr, t)&=&
\int d \bfv m(v_i -u_i)(v_j - u_j)f^{(0)}(\bfr, \bfv, t)\nonumber \\
 &=&  nk_BT(\bfr , t) \delta_{ij}\nonumber \\
 &\equiv &P(\bfr , t) \delta_{ij}, \\
Q_i(\bfr , t)&=&
\int d \bfv \frac{m}{2}\vert \bfv -\bfu\vert ^2 (v_i - u_i)
f^{(0)}(\bfr, \bfv , t)\nonumber \\
&=& 0.
\eeq
We have defined the pressure $P$ using 
 the equation of state for the ideal gas in the second
 line. There is no heat flux because the distribution 
function $f^{(0)}$ in the formulae is the one 
 for  the local equilibrium.
Inserting these formulae into (\ref{balance}),
we end up with a fluid dynamical equation without dissipation,
 i.e., the {\bf Euler equation};
\beq
\frac{\d \rho }{\d t} &+& \nabla \cdot \rho \bfu =0, \\ 
\frac{\d (\rho u_i)}{\d t} &+& \frac{\d}{\d x_j} ( \rho u_i u_j ) + 
\frac{\d}{\d x_i} P = 0, \\
\frac{\d}{\d t}( \rho u^2 + e ) &+& \frac{\d}{\d x_i}
\biggl[ ( \rho \frac{u^2}{2} + e + P ) u_i \biggl] = 0. 
\eeq
We notice that these equations have been obtained from the
 RG equation (\ref{master:0}).
It should be emphasized however that the distribution function 
obtained in the present approximation takes the form
\beq
f(\bfr , \bfv , t)&=&f^{(0)}(\bfr , \bfv , t)\nonumber \\
 &+& A^{-1}{\rm Q}\bfv \cdot \frac{\d f^{(0)}(\bfr , \bfv , t)}{\d \bfr},
\eeq
which incorporates as a perturbation   
a distortion from the local equilibrium distribution and  gives
rise to dissipations.

One can proceed to the second order approximation straightforwardly
and  obtain fluid dynamical equation with dissipations as
 the RG equation.
The perturbation equation in the second order reads
\beq
\label{second}
\biggl(
(\frac{\d}{\d t} - A){\tf}^{(2)}
\biggl)_i 
= - {\bfv}_i \cdot \frac{\d \tf_i^{(1)}}{\d {\tbfr}}.
\eeq
Here, we must make an important notice:
We have actually used the linearized Boltzmann 
equation\cite{resibois}
 neglecting the second-order term of $\tbff ^{(1)}$ in 
the collision integral:
It is known that the neglected term produces  the so called
 Burnett terms which are absent in the usual Navier-Stokes
 equation \cite{landau}.

Applying the RG method, we have
\beq
\frac{\d f^{(0)}}{\d t}&+&\eps {\rm P}\bfv\cdot\nabla f^{(0)}\nonumber \\
 & &+\eps^2 {\rm P}\bfv\cdot \nabla 
A^{-1}{\bf Q}\bfv\cdot \nabla f^{(0)}=0.
\eeq
 The third term  represents dissipations.
Taking inner products between the first equation and
 the particle number, the velocity and
 the kinetic energy, one will obtain a fluid dynamical 
equation with dissipations included, i.e., Navier-Stokes 
equation\cite{resibois}.

\section{Summary and concluding remarks}

In summary, we have shown that the fluid dynamical limit of the 
Boltzmann equation can be obtained  neatly in  the RG method
as formulated in \cite{efk}.
It would be intriguing to see what kind of the dissipative fluid
dynamical would emerge when the RG method is applied to 
the relativistic kinetic equations\cite{muronga}.

The reduction of  a kinetic equation to a further slower
dynamics appear quite often, reflecting the hierarchy of the
space-time of  nature. 
A typical problem 
\cite{fp,brinkman}
in this category
 is to derive Smoluchowski equation\cite{smoluchowski}
 from Kramers
equation\cite{kramers}.
As is expected, it is possible to develop a systematic 
theory for the adiabatic elimination of fast variables in
 Fokker-Planck equations, which appear in the theory of 
the critical dynamics\cite{critical}.

We hope that we can report a development on these problems
in near future.

\end{document}